\documentclass[%
nofootinbib,
 amsmath,amssymb,
 aps,
 prd,
twocolumn, 
superscriptaddress
]{revtex4-1}

\usepackage[normalem]{ulem}
\usepackage{cancel}

\usepackage[dvipsnames]{xcolor}
\usepackage{graphicx}
\usepackage{dcolumn}
\usepackage{bm}
\usepackage{orcidlink}
\hypersetup{colorlinks, citecolor=green}
\hypersetup{colorlinks=true, linkcolor=Bittersweet, urlcolor=MidnightBlue}
\usepackage{hyperref}
\usepackage{ulem}
\usepackage{xcolor}
\usepackage{footnote}

\usepackage{xcolor}
\definecolor{darkblue}{rgb}{0.2, 0.2, 0.6}

\begin{document}


\title{Compact stars in scalar-tensor theories with a single-well potential and the corresponding $f(R)$ theory}

\author{Juan M. Z. Pretel}
 \email{juanzarate@cbpf.br}
 \affiliation{
 Centro Brasileiro de Pesquisas F{\'i}sicas, Rua Dr.~Xavier Sigaud, 150 URCA, Rio de Janeiro CEP 22290-180, RJ, Brazil
}

\author{Sergio E. Jor\'as}
\email{joras@if.ufrj.br}
\affiliation{
 Instituto de F\'\i sica, Universidade Federal do Rio de Janeiro,\\
 CEP 21941-972 Rio de Janeiro, RJ, Brazil 
}

\author{Ribamar R. R. Reis}
 \email{ribamar@if.ufrj.br}
\affiliation{
 Instituto de F\'\i sica, Universidade Federal do Rio de Janeiro,\\
 CEP 21941-972 Rio de Janeiro, RJ, Brazil 
}
\affiliation{Universidade Federal do Rio de Janeiro, Observat\'orio do Valongo, 
 \\CEP 20080-090 Rio de Janeiro, RJ, Brazil}

\author{Sergio B. Duarte}
 \email{sbd@cbpf.br}
 \affiliation{
 Centro Brasileiro de Pesquisas F{\'i}sicas, Rua Dr.~Xavier Sigaud, 150 URCA, Rio de Janeiro CEP 22290-180, RJ, Brazil
}

\author{Jos\'e D. V. Arba\~nil}
\email{jose.arbanil@upn.pe}
\affiliation{Departamento de Ciencias, Universidad Privada del Norte, Avenida el Sol 461 San Juan de Lurigancho, 15434 Lima, Peru}
\affiliation{Facultad de Ciencias F\'isicas, Universidad Nacional Mayor de San Marcos, Avenida Venezuela s/n Cercado de Lima, 15081 Lima, Peru}

\date{\today}

\begin{abstract}
The macroscopic properties of compact stars in modified gravity theories can be significantly different from the general relativistic (GR) predictions. Within the gravitational context of scalar-tensor theories, with a scalar field $\phi$ and coupling function $\Phi(\phi)= \exp[2\phi/\sqrt{3}]$, we investigate the hydrostatic equilibrium structure of neutron stars for the simple potential $V(\phi)= \omega\phi^2/2$ defined in the Einstein frame (EF). From the scalar field in the EF, we also interpret such theories as $f(R)$ gravity in the corresponding Jordan frame (JF). The mass-radius relations, proper mass, and binding energy are obtained for a polytropic equation of state (EoS) in the JF. Our results reveal that the maximum-mass values increase substantially as $\omega$ gets smaller, while the radius and mass decrease in the low-central-density region as we move further away from the pure GR scenario. Furthermore, a cusp is formed when the binding energy is plotted as a function of the proper mass, which indicates the appearance of instability. Specifically, we find that the central-density value where the binding energy is a minimum corresponds precisely to $dM/d\rho_c^J = 0$ on the $M(\rho_c^J)$-curve.

\end{abstract}

\maketitle


\section{Introduction}

Modified theories of gravity have been tested for a long time, in different scenarios, from cosmology to the Solar system. As we will recall in the next section, it is well known that the so-called $f(R)$ theories can be cast as a scalar-tensor one \cite{Sotiriou,DeFelice}, since, upon a suitable conformal transformation, the Lagrangian can be written as the standard Einstein-Hilbert Lagrangian with respect to the new metric, with an extra scalar field subject to a given potential. The original motivation for introducing modifications or extensions to General Relativity (GR) was purely theoretical \cite{Bicknell}. Observations supporting such line of research came only later on, as an option to introducing an unknown component of the Universe. In the beginning one started to search for explanations for the primordial inflationary scenario other than the inflaton field \cite{Starobinsky}. More recently (see, for instance, Ref.~\cite{DeFelice} and references therein), one has started to search for the culprit for the current accelerated expansion of the Universe, other than a plain cosmological constant.

Most of the simplest extensions (in particular, polynomial expressions in $R$) have already been discarded by observational data. Note, however, that some of them do work in limited scenarios, such as the Starobinsky model $f(R) = R + \alpha R^2$ \cite{Starobinsky} --- which is a strong candidate for generating primordial inflation \cite{Planck}, but it is also strongly disfavored for standard cosmology \cite{luca} and stellar stability \cite{us}. The one question is then: is there a $f(R)$ that can replace General Relativity (GR) without spoiling its many achievements and, at the same time, explain the primordial inflation and/or the current accelerated expansion of the Universe and, perhaps, allow extra-massive neutron stars \cite{Demorest_2010} with non-exotic EoSs?  

As a new theory of Gravitation, it must be tested in every possible scenario. In the present work, we will focus on spherically-symmetric relativistic stars, which provide a unique testing environment: a large range of values for the Ricci scalar (from the star core to the radial infinity) and, at the same time, are constrained by a huge amount of data on their masses and radii.  Here, we will focus on the so-called $f(R)$ gravity theories in the metric approach, where $f(R)$ is a yet-to-be-specified non-linear function of the Ricci scalar $R$ and the metric is the fundamental geometric object\footnote{As opposed to the Palatini formalism, which considers both the metric and the connection as independent quantities.}.

Instead of following a systematic search through an infinite-dimensional function space, here we take an alternative approach: we start at the Einstein Frame (EF) --- see a brief review below --- with the simplest possible potential $V(\phi)\sim \phi^2$, perform all the calculations (namely, the numerical solutions to the modified TOV equations) and, at the end of the day, go to the Jordan Frame (JF) --- where the matter energy-momentum tensor is conserved and measures are taken --- finding then the corresponding function $f(R)$. Such approach has been followed before \cite{Peralta} for cosmology, in the inflationary era, providing a powerful analogy to a non-ideal gas that described a phase transition from inflation to a matter-dominated Universe.

In this study, will apply the same approach in relativistic compact stars. Indeed, in the last decade, for instance, isotropic \cite{Mendes2016, Sotani2017} and anisotropic \cite{Silva2015} neutron stars have already been investigated within the framework of massless scalar-tensor theories in the strong-field regime. Through mass-radius diagrams, those works have shown that the spontaneous scalarization phenomenon with massless scalars is capable of leading to significant deviations with respect to the pure GR results. See Ref.~\cite{Stephanie2023} for a recent study on tidal deformability of neutron stars in scalar-tensor theories of gravity in the absence of scalar potential. In addition, Creci \textit{et al.}~\cite{Gaston2023} have demonstrated that three independent tidal deformabilities are needed to characterize the stellar configurations: a scalar, tensor, and a novel ``mixed'' parameter.

However, the observation of the pulsar–white dwarf binary system \cite{Antoniadis2013} has almost ruled out such massless theory \cite{Fethi2016}. Furthermore, if one accounts for the cosmological evolution of the scalar field without the inclusion of a mass term in the action, the original scalar-tensor theory proposed by Damour and Esposito-Farèse is in tension with Solar System constraints \cite{Anderson2016}. Notwithstanding, it has been argued that the incorporation of a massive field to the scalar potential can restore this characteristic of spontaneous scalarization without conflicting with the observations \cite{Fethi2016, Yazadjiev2016, Hu2021, Tuna2022}.

In massive scalar-tensor gravity, the literature offers an extensive study of scalarized compact stars \cite{Yazadjiev2016, Hu2021, Tuna2022, motahar2019, Aguiar_Raissa2006}. As a matter of fact, the relativistic structure of slowly rotating compact stars was studied for three EoSs describing not only hadronic but also strange matter \cite{Staykov2014} (review also \cite{Yazadjiev2014}), and the effects of rapidly rotating on stellar structure and quasinormal modes of neutron stars were respectively investigated in \cite{Yazadjiev_Doneva2015} and \cite{Blazquez_Khoo2021, Blazquez_gonzalez2022} in a scalar-tensor theory, where the scalar potential is associated with the aforementioned $R$-squared gravity. In this case, the EF potential corresponds to a massive scalar field with a mass $m_\phi= 1/\sqrt{6\alpha}$. Following a similar procedure, Staykov \textit{et al.}~\cite{Staykov2016PRD} determined the crust-to-core ratio of the moment of inertia of neutron stars in the same theory. See also Refs.~\cite{Yazadjiev2016} and \cite{Doneva2016JCAP} for the respective study of slowly and rapidly rotating neutron stars in scalar-tensor theories with a simple dilaton potential $V(\phi)= m_\phi^2\phi^2 /2$ yielding the massive scalar field $\phi$. Specifically, the authors showed that the inclusion of a mass term for the scalar field indeed changes the picture drastically compared to the massless case.

Other authors discussed the structure of the weakly and strongly scalarized branches of neutron star models in two-derivative scalar-tensor theories of gravity with a massive scalar field \cite{sym12091384}. The tidal deformability and X-ray pulse profiles of scalarized neutron stars were recently analyzed for three different types of massive scalar-tensor theories \cite{Hu2021}. It is worth commenting that neutron stars in scalar-tensor theories with a quartic, symmetry-breaking potential (also known as the symmetron model) have recently been explored \cite{Aguiar2022}. Neutron star phenomenology in inflationary attractor scalar-tensor theories can be found in Refs.~\cite{OikonomouMNRAS2023, OikonomouCQG2023, Odintsov2023}. Phenomenological features and the stability of boson stars in massless and massive scalar-tensor gravity were recently addressed \cite{Evstafyeva2023}. In addition, for a comprehensive discussion on scalarization and its manifestation in compact objects, we refer the reader to the recent review article \cite{Doneva2022Review}.

Motivated by the several studies already mentioned, here we investigate the relativistic structure of neutron stars in scalar-tensor theories with coupling function $\Phi(\phi)= \exp[2\phi/\sqrt{3}]$ for a single-well potential $V(\phi)$ defined in the EF. In addition to calculating the macroscopic properties of the equilibrium configurations, we further interpret such theories as $f(R)$ gravity in the corresponding JF.

Actually, that is the main difference between the present work and the previous literature \cite{Staykov2014, Yazadjiev2014, Yazadjiev_Doneva2015, Blazquez_Khoo2021, Blazquez_gonzalez2022, Staykov2016PRD}: by taking that final step to a $f(R)$ theory, we rely on a stronger motivation for an otherwise {\it ad-hoc} choice of the function $f(R)$. Instead of assuming a particular function, we assume a simple and robust single-well massive potential for the scalar field $\phi$ in the Einstein frame. Here, as before \cite{Peralta}, the function $f(R)$ is indeed multi-valued, but we focus on the almost-linear branch. The reason is threefold: first, we do not expect that an actual star probes the unstable (middle) branch (where $f''<0$, see details below), for it would render the whole star unstable and therefore, non-existing for observational purposes. Secondly, the last branch, being almost linear, is closer to GR, i.e., we expect it will not spoil the many successes of GR, while being still able to be distinguishable from it (since $f''(R)\neq 0$). Thirdly, the $f(R)$ theory we arrive at is {\it not} Starobinsky's, since its second derivative is not constant and does vanish dynamically, by itself, as the scalar field oscillates around its minimum. Therefore, our approach is indeed orthogonal to the ones already pursued in the literature (both in stellar and in cosmological backgrounds) and yields different results (although, in the present case, numerically similar to those of GR, as discussed above).

The organization of the present work is therefore as follows: In Sec.~\ref{framesSec} we begin by defining the EF action and the corresponding functions $f(R)$ and $R$ in the JF. We then present the modified TOV equations and define the gravitational binding energy for polytropic compact stars in Sec.~\ref{TOVEqsSec}. Section \ref{ResultsSec} provides a discussion of the numerical results in terms of the mass-radius diagram, function $f(R)$, and binding energy. We conclude with some closing comments in Sec.~\ref{Conclusions}.

\section{Frames}\label{framesSec}

All quantities below are defined either in the Jordan Frame (JF) or in the Einstein Frame (EF) by a sub(super)script ``J" or ``E", respectively.

The starting point is the modified gravity Lagrangian in the JF:
\begin{equation}
{\cal L}_J \equiv \sqrt{-g_J} f(R_J),
\end{equation}
to which one can add the Lagrangian for any matter/radiation field that is also present in the Universe. The standard approach is to perform a Legendre transformation so that the independent variable $R$ is replaced by $\Phi \equiv f'\equiv df/dR$:
\begin{equation}
{\cal L}_J = \sqrt{-g_J} \bigg[ \Phi R(\Phi) - W(\Phi)\bigg],
\label{legendre}
\end{equation}
where $ W(\Phi) \equiv \Phi R(\Phi) - f[R(\Phi)]$ and $R(\Phi)$ is obtained from inverting the very definition of $\Phi$. For that, we require that $f''(R) \equiv d^2f/dR^2 \neq 0$, expect, perhaps, only at particular values. Note that Eq.~(\ref{legendre}) is {\it not} a Brans-Dicke Lagrangian and that the field $\Phi$, in spite of no explicit kinetic term, does have an equation of motion of its own. It precisely expresses the extra degree of freedom present in $f(R)$ theories.

One then performs a conformal transformation to the EF
\begin{equation} 
g_{\mu\nu}^E = 
\Phi g_{\mu\nu}^J.
\end{equation}
Since $\Phi>0$, one can express it as a Real (dimensionless) field $\phi\equiv \sqrt{3}/2 \ln\Phi$, which leads us to the Action in the EF:
\begin{align}
    S =&\ \frac{1}{2\kappa}\int d^4x\sqrt{-g_E}\left[ R_E- 2g^{\mu\nu}_E\partial_\mu\phi \partial_\nu\phi-  V(\phi) \right] + \nonumber  \\
    &+ S_m\left( 
    \Phi^{-1} g_{\mu\nu}^E, \Psi \right) , 
\label{ActionEq}
\end{align}
where $\kappa= 8\pi$, $R_E$ is the Ricci scalar with respect to the Einstein frame metric $g_{\mu\nu}^E$ (whose determinant is $g_E$), and $S_m$ is the action of  matter/radiation fields collectively denoted by $\Psi$.  The fields $\Psi$ will still follow geodesics given by the original metric $g^J_{\mu\nu}$; its energy-momentum tensor is {\it not} conserved using $g^E_{\mu\nu}$ due to the explicit non-minimal coupling to $\Phi$. Therefore, we consider the former metric as the physical one. Of course, in GR, $\Phi\equiv f' = 1$ and there is no difference between the frames.

\begin{figure}
\center
\includegraphics[width=0.47\textwidth]{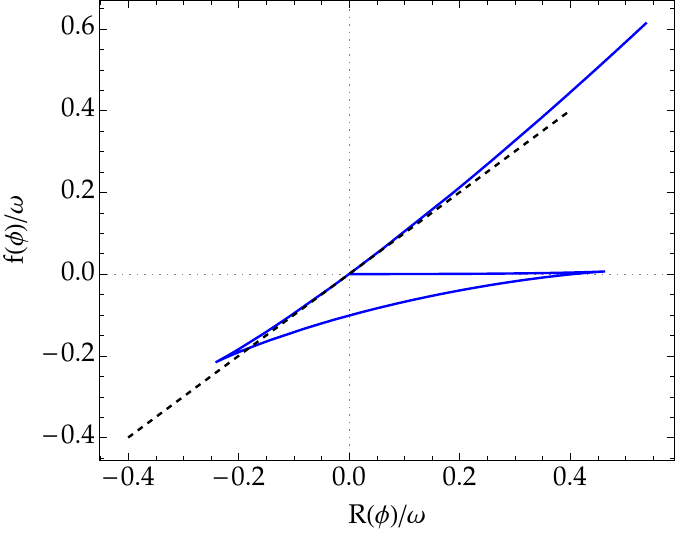}
\caption{Behavior of $f(R)$ that corresponds to a single-well $V(\phi)$ in the EF. The bottom of the well (at $\phi=a$) corresponds to $R=0$, around which the field $\phi$ and $R$ oscillate. The first branch, almost horizontal, has $f'\ll 1$ but also  $f''>0$. The straight dashed black line is GR, i.e, $f(R) =R$. We have used $a=0$, but different values of $a$ would only change the ranges on both axes; the qualitative picture, namely the 3-branch structure, would remain.}
\label{fR}
\end{figure}

As we mentioned above, here we reverse the usual approach and start at the EF, where assume the simplest non-trivial functional form for $V(\phi)$ in the EF, namely
\begin{equation}
V(\phi) = \frac{1}{2} \omega 
(\phi-a)^2.
\label{Vphi}
\end{equation}
Following a previously established procedure \cite{Magnano:1993bd}, one is able to define the corresponding $f(R)$ in the JF, in parametric form:
\begin{align}
\label{fphi}
f(\phi) &= {\rm e}^{4 \phi/\sqrt{3}} \left[2V(\phi) +\sqrt{3}  \frac{{\rm d} V(\phi)}{d\phi} \right], \\
\label{Rphi}
R(\phi) &= {\rm e}^{2 \phi/\sqrt{3}} \left[4 V(\phi) +\sqrt{3} \frac{{\rm d} V(\phi)}{d\phi} \right],
\end{align}
where\footnote{Since $\phi$ is dimensionless, $\omega$ and $R$ are given in units of $M_{pl}^2$.}, plotted in Fig.~\ref{fR}, from where we can take two important pieces of information: (i) $f' (R)>0 \, \forall R$, as expected, and (ii) $f'' <0$ only in the lower branch. Since the latter indicates instabilities, we do not expect that this range is probed in the present scenario, except, perhaps, on standard unstable branches in mass-radius diagrams (see, however, discussion in the final section). It is worth mentioning that the bottom of the potential (\ref{Vphi}), i.e $\phi=a$, corresponds to $f=0=R$ and that the full range of $R$ around that final value growths exponentially with $\phi$.  Since a non-vanishing value for $a$ would yield only scaling differences in the outcome, we will assume $a=0$ from now on.

Taking the variation of the above action (\ref{ActionEq}) with respect to the metric tensor $g_E^{\mu\nu}$ and the scalar field $\phi$, we can obtain the field equations in the Einstein frame, respectively, 
\begin{align}
    &G_{\mu\nu}^E = \kappa \left( T_{\mu\nu}^E + T_{\mu\nu}^\phi \right),  
    \label{FieldEq1}  \\
    &\square^{E}\phi- \frac{1}{4}\frac{dV(\phi)}{d\phi} = \frac{\kappa}{2\sqrt{3}}
    T_E ,  \label{FieldEq2}
\end{align}
where 
\begin{equation}
  \kappa T_{\mu\nu}^\phi\equiv 2\partial_\mu\phi \partial_\nu\phi - g_{\mu\nu}^E g_E^{\alpha\beta}\partial_\alpha\phi \partial_\beta\phi - \frac{1}{2}V(\phi)g_{\mu\nu}^E 
\end{equation}
is the standard energy-momentum tensor of the scalar field $\phi$ and $\square^{E} = \nabla^{E}_\mu\nabla_{E}^\mu$ denotes the d'Alembert operator in curved spacetime with $\nabla^{E}_\mu$ standing for the covariant derivative in the Einstein frame, i.e., calculated from $g^{E}_{\mu\nu}$. Moreover, $T_E \equiv g_E^{\mu\nu}T_{\mu\nu}^E$ is the trace of the energy-momentum tensor
\begin{equation}
    T_{\mu\nu}^E = \frac{-2}{\sqrt{-g_E}}\frac{\delta S_m}{\delta g_E^{\mu\nu}}.
\end{equation}

The four-divergence of Eq.~(\ref{FieldEq1}) leads to 
\begin{equation}
    \kappa\nabla^\mu_E T_{\mu\nu}^E + 2\nabla_\nu^E \phi \square^E\phi - \frac{1}{2}\nabla_\nu^E V(\phi) =0 ,
\end{equation}
and, in view of Eq.~(\ref{FieldEq2}), we obtain the conservation law for the Einstein frame energy-momentum tensor 
\begin{equation}\label{ConservEq}
    \nabla^\mu_E T_{\mu\nu}^E= - \frac{1}{\sqrt{3}} T_E\nabla_\nu^E\phi .
\end{equation}
As expected, only massless fields, whose trace of the energy-momentum tensor vanishes, are covariantly conserved both before and after a conformal transformation (i.e., using either metric).

We consider spherically symmetric and static metrics as follows: 
\begin{align}
    ds_E^2 &= -{\rm e}^{2\nu_E(r_E)}dt^2 + {\rm e}^{2\lambda_E(r_E)}dr_E^2 + r_E^2d\Omega^2 ,  \\
    ds_J^2 &=  -{\rm e}^{2\nu_J(r_J)}dt^2 + {\rm e}^{2\lambda_J(r_J)}dr_J^2 + r_J^2d\Omega^2 ,
\end{align}
where $d\Omega^2= d\theta^2+ \sin^2\theta \,d\varphi^2$ is the line element on the unit 2-sphere --- we recall the reader that conformal transformations do not affect angles. Consequently, from $ds_E^2= \Phi \, ds_J^2$, we obtain the following relations between both frames:
\begin{subequations}
\begin{align}
    r_E^2 &= \Phi\, r_J^2,\label{rEJ} \\
    {\rm e}^{2\nu_E} &= \Phi \,{\rm e}^{2\nu_J}, \label{MetricRelatFramesEq1}  \\
    {\rm e}^{-2\lambda_J} &= \Phi \,{\rm e}^{-2\lambda_E}\left[ \frac{dr_J}{dr_E} \right]^2 \nonumber \\
    & = {\rm e}^{-2\lambda_E} \left[ 1- \frac{r_E\phi'(r_E)}{\sqrt{3}} \right]^2 . \label{MetricRelatFramesEq2}
\end{align}
\end{subequations}
The first relation (\ref{rEJ}) means that the physical radial coordinate as measured in the Jordan frame is $r_J = r_E/\sqrt{\Phi}$, and hence the physical radius of the compact star will be given by $r_{\rm sur}^J= \exp(-\phi_{\rm sur}/\sqrt{3})
r_{\rm sur}^E$. Meanwhile, the expression (\ref{MetricRelatFramesEq1}) implies that
\begin{equation}
    \nu_J[r_J(r_E)] = \nu_E(r_E) - \frac{\phi(r_E)}{\sqrt{3}} .
\end{equation}

Furthermore, in the JF, we can define the mass function by means of the following relation
\begin{equation}
    {\rm e}^{-2\lambda_J(r_J)} = 1- \frac{2m_J(r_J)}{r_J}.
\end{equation}
Thus,  the physical mass function can be written in terms of the Einstein-frame quantities as
\begin{align}\label{JordanMass}
    m_J[r_J(r_E)] =&\ {\rm e}^{-\phi/\sqrt{3}} \frac{r_E} {2}  \times \nonumber \\
    & \times \left[ 1- {\rm e}^{-2\lambda_E}\left( 1- \frac{r_E\phi'(r_E)}{\sqrt{3}} \right)^2 \right]  ,
\end{align}
from which it is evident that the masses in both frames coincide at sufficiently large distances. For this reason, the asymptotic mass (i.e., measured at infinity) of a compact star will simply be denoted by the letter $M$. \\

\section{Modified TOV equations and binding energy}\label{TOVEqsSec}

To describe the dense-matter source, we consider that the compact star is composed of an isotropic perfect fluid, whose energy-momentum tensor in the Einstein frame is given by
\begin{equation}
    T_{\mu\nu}^E = (\rho_E+ p_E)u_\mu^E u_\nu^E + p_Eg_{\mu\nu}^E ,
\end{equation}
where $\rho_E$, $p_E$, and $u_\mu^E$ are the energy density, pressure, and four-velocity of the fluid in the Einstein frame, respectively. The EF energy-momentum tensor $T_{\mu\nu}^E$ is related to the JF one via $T_{\mu\nu}^E = \Phi^{-1}
T_{\mu\nu}^J$, so that we can write 
\begin{subequations}\label{ThermoEJframes}
\begin{align}
    \rho_E &= \Phi^{-2}
    \rho_J,\\
    p_E &= \Phi^{-2}
    p_J,  \\
    u_\mu^E  &= \Phi^{1/2}
    u_\mu^J .
\end{align}
\end{subequations}

In the hydrostatic equilibrium state (i.e., when both the metric and thermodynamic variables do not depend on the time coordinate), we have $u_E^\mu= ({\rm e}^{-\nu_E}, 0, 0, 0)$ and $T_{\mu\nu}^E= {\rm diag} ({\rm e}^{2\nu_E}\rho_E, {\rm e}^{2\lambda_E}p_E, r_E^2p_E, r_E^2p_E\sin^2\theta)$. Consequently, the field equations (\ref{FieldEq1}) and (\ref{FieldEq2}), together with the conservation equation (\ref{ConservEq}) lead to the following set of differential equations in the Einstein frame

\begin{widetext}
\begin{subequations}
  \begin{align}
      \frac{1}{r_E^2}\frac{d}{dr_E}\left[ r_E(1- {\rm e}^{-2\lambda_E}) \right] &= \kappa\, \rho_E + \frac{1}{{\rm e}^{2\lambda_E}}\left[ \frac{d\phi}{dr_E} \right]^2 + \frac{1}{2}V(\phi) ,  \label{TOV1}  \\
      \frac{2}{r_E}{\rm e}^{-2\lambda_E}\frac{d\nu_E}{dr_E} - \frac{1}{r_E^2}\left(1- {\rm e}^{-2\lambda_E}\right) &= \kappa \, p_E+ \frac{1}{{\rm e}^{2\lambda_E}}\left[ \frac{d\phi}{dr_E} \right]^2 - \frac{1}{2}V(\phi) ,  \label{TOV2}  \\
      \frac{d^2\phi}{dr_E^2}+ \left[ \frac{d\nu_E}{dr_E} - \frac{d\lambda_E}{dr_E} + \frac{2}{r_E} \right]\frac{d\phi}{dr_E} &= -\frac{\kappa}{2\sqrt{3}}(\rho_E - 3p_E){\rm e}^{2\lambda_E} + \frac{1}{4}\frac{dV(\phi)}{d\phi} {\rm e}^{2\lambda_E} ,   \label{TOV3}  \\
      \frac{dp_E}{dr_E} &= -(\rho_E+ p_E)\left[ \frac{d\nu_E}{dr_E} - \frac{1}{\sqrt{3}}\frac{d\phi}{dr_E} \right] - \frac{4p_E}{\sqrt{3}}\frac{d\phi}{dr_E} .  \label{TOV4}
  \end{align}
\end{subequations}
\end{widetext}

The relations for energy density and pressure in both frames (\ref{ThermoEJframes}) are often substituted into the above stellar structure equations so that the numerical solution provides the thermodynamic quantities in the physical Jordan frame \cite{Yazadjiev2014, Staykov2014, Yazadjiev2016}. In our numerical calculations we will follow the same procedure but considering a polytropic EoS in the JF, namely
\begin{equation}\label{EquaoS}
    p_J = K_J\rho^\gamma_J ,  
\end{equation}
where $\gamma \equiv 1+1/n$ and $n$ is the polytropic index. Its analytical simplicity has allowed the description of neutron stars within the context of both GR \cite{Allen1998, KokkotasRuoff2001} and modified gravity  \cite{Pretel2021}. If we recall that the relation between the energy densities and pressures in each frame is
\begin{align}
\rho_E(r) &\equiv \Phi^{-2}[\phi(r)] \, \rho_J(r),
\label{rhophi}\\
p_E(r) &\equiv \Phi^{-2}[\phi(r)] \, p_J(r),
\label{pphi}
\end{align}
we then notice that the equation of state in the JF can be written in terms of EF quantities as
\begin{align}
p_E &= (K_J \Phi^{2(\gamma-1)}) \rho_E^\gamma\nonumber\\
p_E &= K_E \, \rho_E^\gamma.
\end{align}
In other words, in the EF, the baryonic matter still is polytropic with the same index $\gamma$ but a different coefficient $K_E[\phi(r)]\equiv K_J \Phi^{2(\gamma-1)}$. We notice, however, that $K_E$ is {\it not} a constant. Therefore, the pressure cannot be written solely as a function of the energy density, namely, the fluid is {\it not} barotropic and can present entropy perturbations in the EF. 

In the numerical results below, we will stick to $\gamma =2$ ($n=1$) and $K_J= 10^8\, \rm m^2$, in geometric units.
The Einstein frame radius of the star $r_{\rm sur}^E$ is reached when the pressure vanishes, i.e.~$p_J(r_{\rm sur}^E)= 0$. Thus, the physical radius of the star will be determined from the relation
\begin{equation}
    r_{\rm sur}^J = r_{\rm sur}^E \Phi^{-1/2}(\phi_{\rm sur}),
\end{equation}
and the total gravitational mass of the star is calculated from the expression (\ref{JordanMass}) evaluated at infinity, given by $M \equiv m_J(r_{J} \rightarrow \infty)$. It should be noted that the Jordan and Einstein frame masses coincide at radial infinity since the scalar field drops off exponentially fast to zero and, therefore,  $\Phi\to 1$.

To numerically solve the modified TOV equations (\ref{TOV1})-(\ref{TOV4}), we follow the standard procedure, that we briefly skecth below for the sake of completeness. We integrate from the center to the surface of the compact star using the following initial values
\begin{align}\label{BoundCondi}
    \rho_J(0) &= \rho_c^J,  &  \nu_E(0) &= \nu_c^E,  &  \lambda_E(0) &= 0,  \nonumber   \\
    \phi(0) &= \phi_c,  &  \phi'(0) &= 0 ,
\end{align}
where $\rho_c^J$ and $\phi_c$ are central values of the energy density and the scalar field, respectively. We then integrate the system of equations from the surface to a sufficiently far distance satisfying the asymptotic flatness requirement:
\begin{align}\label{AsympFlatRequi}
    \lim_{r\rightarrow \infty} \phi (r) &= 0,  &  \lim_{r\rightarrow \infty} m (r) &= {\rm constant} .
\end{align}
Specifically, $\phi_c$ must be properly chosen in order to obey (\ref{AsympFlatRequi}) at great distances from the stellar surface. Furthermore, the central value of the metric function $\nu_E$ in Eq.~(\ref{BoundCondi}) is chosen so that $\nu_E(r\rightarrow\infty) \rightarrow 0$.

If one writes Eqs.~(\ref{TOV1}) and (\ref{TOV2}) as, respectively,
\begin{subequations}
\begin{align}
\frac{1}{r_E^2} \left(1 - {\rm e}^{-2 \lambda_E}\right) + 
  \frac{2}{r_E} {\rm e}^{-2 \lambda_E}\lambda'_E &= \kappa [ \rho_E + \rho_\phi],
  \label{tov2b}\\
\frac{2}{r_E}{\rm e}^{-2 \lambda_E} \nu'_E - 
  \frac{1}{r_E^2} \left(1 - {\rm e}^{-2 \lambda_E}\right) &=
 \kappa  [p_E + p_\phi],
  \label{tov1b}
\end{align}
\end{subequations}
the standard expressions for the energy density and the pressure of the scalar field $\phi(r)$ can be promptly identified:
\begin{align}
\kappa\,\rho_\phi &\equiv 
  {\rm e}^{-2 \lambda_E}\phi'^2 + \frac{1}{2}\,V(\phi),\\
\kappa\, p_\phi &\equiv {\rm e}^{-2 \lambda_E} \phi'^2 - \frac{1}{2} V(\phi) .
\end{align}

An important quantity that serves as an indicator of the instability of compact stars is the ``gravitational binding energy'', which is negative for a gravitationally bound system and it is defined as the difference
\begin{equation}\label{BEnergEq}
    E_B = M - M_{\rm pr} , 
\end{equation}
where $M_{\rm pr}$ is the proper mass, also known in the literature as baryon mass since it is defined by the volume integral of the baryon number density times the mass of the baryons \cite{Alecian2004}. In other words, the JF proper mass is given by 
\begin{equation}\label{PropMass1}
    M_{\rm pr}^J = \int \rho_{\rm rest}^J dV_{\rm pr}^J = 4\pi\int_0^{r_{\rm sur}^J} \rho_{\rm rest}^J {\rm e}^{\lambda_J} r_J^2dr_J ,
\end{equation}
with $\rho_{\rm rest}^J$ and $dV_{\rm pr}^J= \sqrt{^{(3)}g_J}d^3x_J = {\rm e}^{\lambda_J} r_J^2\sin\theta dr_Jd\theta d\varphi$ being the rest mass density and the proper volume element in the physical Jordan frame, respectively. Through the relations (\ref{rEJ}) and (\ref{MetricRelatFramesEq2}), the above expression becomes
\begin{align}
    M_{\rm pr}^J &= 4\pi\int_0^{r_{\rm sur}^E} \Phi^{-3/2}[\phi(r_E)] \rho_{\rm rest}^J {\rm e}^{\lambda_E} r_E^2dr_E 
 \nonumber  \\
    &= 4\pi\int_0^{r_{\rm sur}^E} \Phi^{-3/2}[\phi(r_E)] 
    \frac{{\rm e}^{\lambda_E} r_E^2 \rho_J}{1+ K_J\rho_J}dr_E , \label{PropMass2}
\end{align}
where we have used the fact that for a usual polytropic EoS we can write the rest mass density as $\rho^J_{\rm rest} = \rho_J/({1+ K_J\rho_J})$ for $\gamma= 2$ ~\cite{Pretel2021} .

In what follows, we will assume a single-well potential (or simple dilaton potential) given by Eq.~(\ref{Vphi}) with $a=0$, which represents a scalar field with constant mass $m_\phi= \sqrt{\omega}$. In our numerical calculations we are going to consider two values for $\omega$, given in units of $r_g^{-2}$, where $r_g= GM_\odot/c^2 \approx 1.477\, \rm km$ is the solar mass in geometric units. Notice that $\omega= 1.0 r_g^{-2}$ corresponds to a mass of the scalar field $m_\phi \sim 10^{-38} M_{pl}$, where $M_{pl}$ is the Planck mass in natural units.

\section{Numerical results}\label{ResultsSec}

\begin{figure}
    \includegraphics[width=0.47\textwidth]{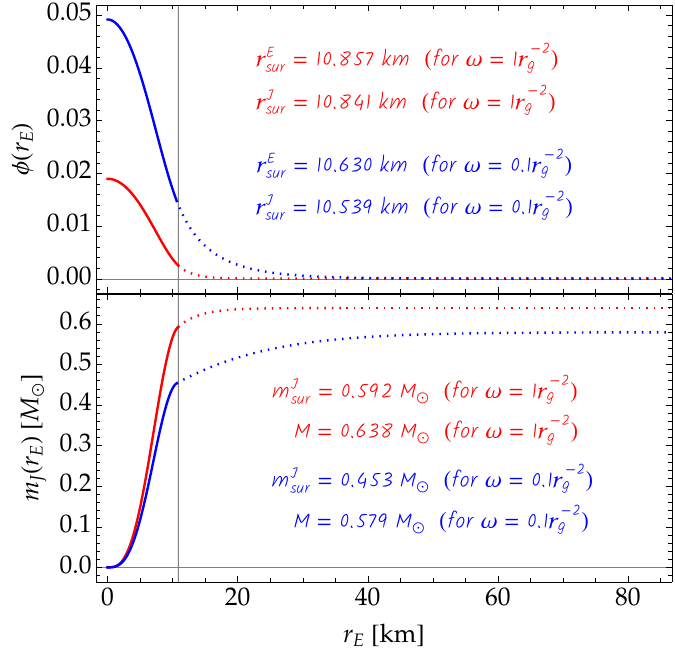}
       \caption{Numerical solution of the modified TOV equations (\ref{TOV1})-(\ref{TOV4}) for a given central density $\rho_c= 0.8 \times 10^{18}\, \rm kg/m^3$ for $\omega= 1.0 r_g^{-2}$ (red curves) and $\omega= 0.1 r_g^{-2}$ (blue curves). The scalar field (top panel) and the mass function in the physical frame (lower panel) are shown as a function of the radial coordinate in the EF. The solid and dotted lines correspond to the interior and exterior solutions, respectively, while the vertical lines indicate the surface radius of the star in the EF. }
    \label{FigNumeSol}
\end{figure}

Using the boundary conditions (\ref{BoundCondi}) and (\ref{AsympFlatRequi}) we numerically solve  the modified TOV equations (\ref{TOV1})-(\ref{TOV4}) inside and outside the compact star. First, we integrate from the center at $r_E= 0$ up to the radius of the star at $r_E= r_{\rm sur}^E$, which is determined by the condition that the pressure vanishes at the surface. Meanwhile, for the exterior problem, we integrate from the surface up to a far enough distance where the asymptotic value of the scalar field is zero, which is due to the particular form of the scalar potential. In other words, the central value of the scalar field $\phi_c$ is chosen such that $\phi(r_E \rightarrow \infty) \rightarrow 0$ given a specific stellar configuration. For instance, for a central density $\rho_c= 0.8 \times 10^{18}\, \rm kg/m^3$, we display in Fig.~\ref{FigNumeSol} the radial profile for the scalar field (top panel) as well as the radial behavior for the JF mass function (bottom panel) which stabilizes to the so-called astrophysical mass as we move away from the stellar surface. It can be observed that the radius, surface mass, and asymptotic mass in the physical frame decrease as $\omega$ becomes smaller.
A smaller $\omega$ has a larger ``leak" of energy (or the so-called curvature fluid) out of the star. It can be explained by two different mechanisms: (i) a lighter field does propagate further away from the star (in the absence of the chameleon effect, which is precisely the case at hand) or (ii) a potential $V(\phi$) with a smaller second derivative corresponds to a weaker restorative force, which then yields a longer path to the ground state at $\phi=0$, i.e, GR.

\begin{figure}
    \centering
    \includegraphics[width=0.46\textwidth]{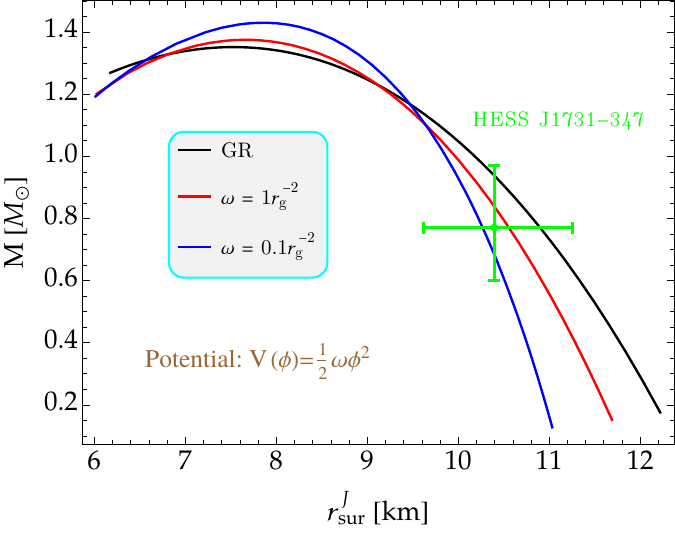}
       \caption{Mass-radius diagram for compact stars with polytropic EoS for two different values of $\omega$. The pure general relativistic solution has been included as a benchmark by the black line. The decrease in the free parameter $\omega$ implies larger maximum-mass values with respect to the GR counterpart. The green dot with its corresponding error bars stands for the central compact object within the supernova remnant HESS J1731-347 \cite{Doroshenko2022}. }
    \label{FigMassRadius}
\end{figure}

\begin{figure}
    \centering
    \includegraphics[width=0.45\textwidth]{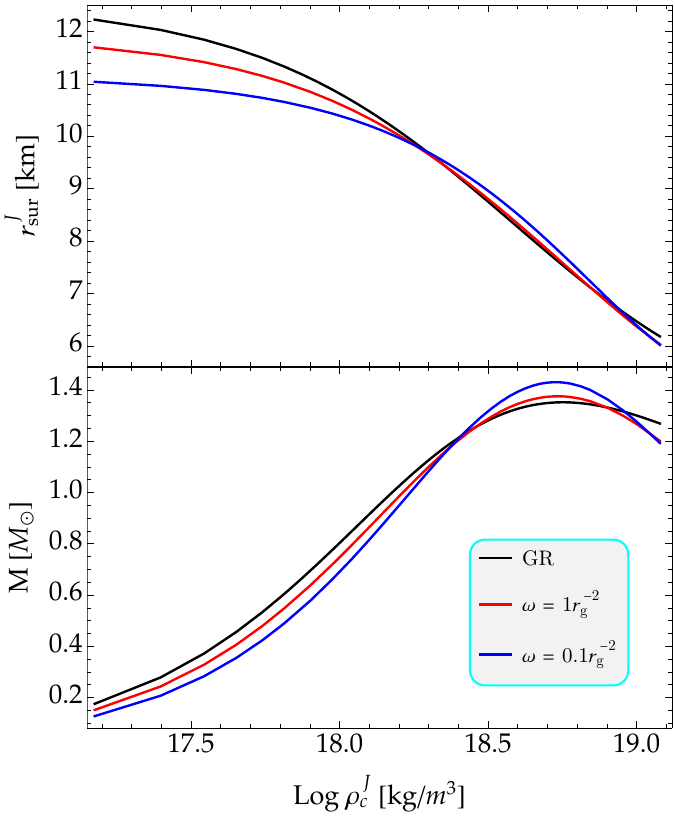}
       \caption{Radius (upper panel) and mass (lower panel) as functions of the central energy density for the equilibrium configurations shown in Fig.~\ref{FigMassRadius}. In the low-central-density branch, both the radius and the mass decrease as we move away from the pure GR context, while this behavior is reversed after a certain value of $\rho_c^J$. }
    \label{FigMassCentDend}
\end{figure}

\begin{figure}
    \centering
    \includegraphics[width=0.465\textwidth]{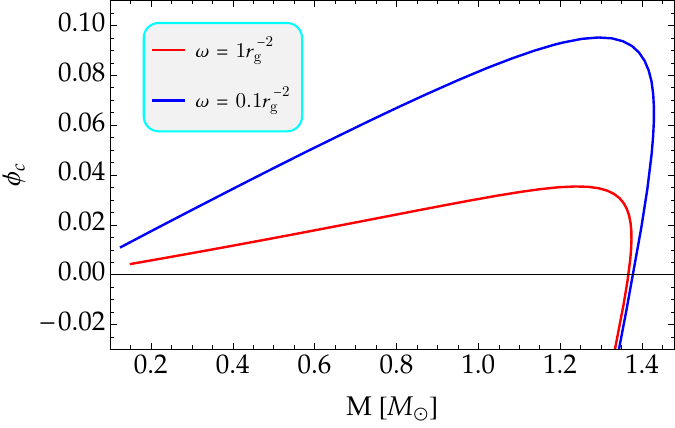} 
       \caption{Scalar field at the stellar center versus the asymptotic mass of compact stars. For each central density, these values of $\phi_c$ have been chosen in such a way that they satisfy the asymptotic flatness requirement (\ref{AsympFlatRequi}). }
    \label{FigPhiCentral}
\end{figure}

As a result of varying the central density $\rho_c$, we obtain the well-known mass-radius diagram. Figure \ref{FigMassRadius} illustrates the total mass as a function of radius for two values of the $\phi$-field mass $\omega$, where we can see that the maximum-mass values undergo a significant increase as $\omega$ decreases. Furthermore, with the appearance of the scalar field, the radius of the star becomes smaller in the low-mass region, however, such behavior can be reversed after a certain value of central density. This can be better appreciated in Fig.~\ref{FigMassCentDend} for low central densities, where both maximum mass and radii change by a $\sim 10\%$ factor with a 10-fold drop in $\omega$. In Fig.~\ref{FigPhiCentral} we show the values of the scalar field at the center of the star versus the gravitational mass for the two choices of $\omega$. Before reaching maximum mass, the central scalar field decreases with increasing $\omega$. This is consistent with the mass-radius relation since for a certain value of $\omega> 1.0 r_g^2$ we recover the GR solution where the scalar field $\phi \rightarrow 0$. Notice that, in the GR limit, we must have $V(\phi)= 0$ and $\Phi= 1$ (or alternatively $\phi= 0$).


Doroshenko \textit{et al.}~\cite{Doroshenko2022} have analyzed the central compact object within the supernova remnant HESS J1731-347, and they estimated the mass and radius of such an object to be $M= 0.77_{-0.17}^{+0.20}\, M_\odot$ and $r_{\rm sur}= 10.4_{-0.78}^{+0.86}\, \rm km$, respectively. This estimate has also been included by the green bars in Fig.~\ref{FigMassRadius}. Our mass-radius results show that the remnant HESS J1731-347 can be described in massive scalar-tensor theories (for the adopted values of $\omega$) as a neutron star with polytropic EoS (\ref{EquaoS}). We must point out that, although the EoS considered in this work generates relatively low maximum masses (compared to other more realistic EoSs), the effect of the scalar field and consequently the qualitative behavior in the mass-radius diagram is maintained for other hadronic matter EoSs.

\begin{figure*}
    \centering
    \includegraphics[width=0.45\textwidth]{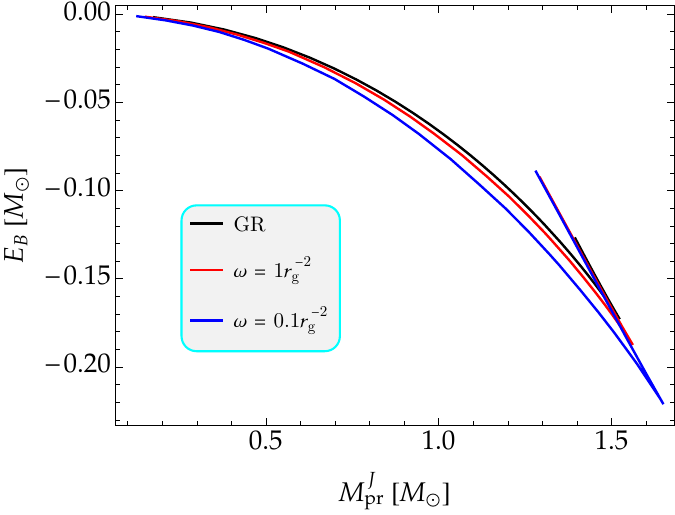} 
    \includegraphics[width=0.452\textwidth]{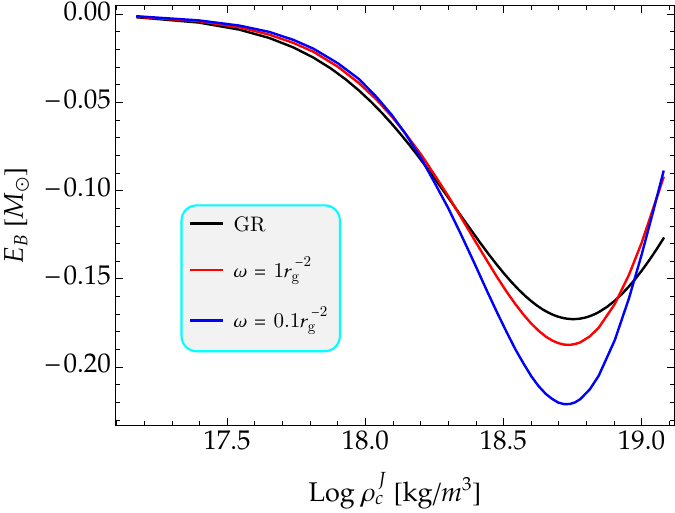}
       \caption{Gravitational binding energy as a function of the proper mass (left panel) and of the central mass density (right panel) for the compact stars displayed in Fig.~\ref{FigMassRadius}. The appearance of a cusp can be observed when the binding energy is minimal for the two values of $\omega$. Such minima correspond precisely to the central densities where $dM/d\rho_c^J = 0$  on the $M(\rho_c^J)$-curves (see the bottom plot of Fig.~\ref{FigMassCentDend}). }
    \label{FigBindEnergy}
\end{figure*}

As usual in GR, the formation of a cusp (when the binding energy is a minimum) could also be used as an indicator to establish the onset of instability in modified gravity theories. With that in mind, in Fig.~\ref{FigBindEnergy} we plot the gravitational binding energy (\ref{BEnergEq}) as a function of proper mass (\ref{PropMass2}) and central density $\rho_c^J$, where the emergence of a cusp in the high-mass region for the adopted massive scalar-tensor theory can clearly be observed. Furthermore, we emphasize that the minima in each curve of the right plot correspond precisely to a certain central-density value where the mass is maximum according to the lower panel of Fig.~\ref{FigMassCentDend}. Therefore, within the gravitational framework of scalar-tensor theories with a simple dilaton potential, the concept of binding energy is important when constructing stable equilibrium configurations since it is compatible with the standard stability criterion in Einstein gravity $dM/d\rho_c^J = 0$.


\section{Conclusions}\label{Conclusions}

 The only solutions we were able to find were close to the GR ones, i.e., in the almost-linear branch in Fig.~\ref{fR}. Therefore, it is expected that they all feature values for both the central densities and total masses similar to those of GR. 

Solutions that cross the 3 branches in Fig.~\ref{fR} are not possible: even a small layer in the unstable region ($f''<0$) jeopardizes the whole star. It is important to point out that this instability has a different nature from the one in GR. Here, even in the standard unstable branch of the mass-radius plot, we have $f''>0$. That does not prevent, however, the existence of non-trivial solutions belonging solely to the first branch (the one with $\Phi\equiv f'\ll 1$ in Fig.~\ref{fR}). It is clear from Eq.~(\ref{rEJ}) that, in this case, $r^J_{\rm sur}\gg r^E_{\rm sur}$,  which would probably lead to unusually low central densities. We point out that, in that case, there would be a larger effective Newton's Constant $G_N^{\rm eff}\equiv G_N/\Phi$ inside and in the immediate neighborhood of the star (we recall that $\phi\to 0$ and $\Phi\to 1$ at radial infinity). Finding such solutions (or discarding their existence) would require a fine-grid search in the parameter space, taking into account also the variation of an extra parameter: a shift in the ground state of $\phi$, in order to allow for a correspondingly large variation on $R$. The search for such solutions is beyond the scope of the current paper.

We also point out a qualitative change in $f(R)$, not investigated here: the unstable branch (where $f ''<0$) is known to be absent if one adds an {\it ad-hoc} $\Lambda$ to Eq.~(\ref{Vphi}),  whose value is above a critical number \cite{Peralta}. Nevertheless, reasonable values of $\Lambda$ (i.e., within  observational constraints) are not expected to have any measurable impact on the star's structure, mass, or radius. Even so, such non-trivial solutions also may be worth exploring in a future work.

For the adopted values of $\omega$, we have found that the maximum mass on the $M(\rho_c^J)$-curve increases with the decrease of the mass of the scalar field $m_\phi$, while the radius and mass decrease in the low-central-density branch. In addition, similar to the GR case, a cusp is formed when the gravitational binding energy is plotted as a function of the proper mass, which indicates the appearance of stellar instability within the framework of massive scalar-tensor theories with a simple dilaton potential. In other words, our results revealed that the central-density value where the binding energy is a minimum corresponds to $dM/d\rho_c^J = 0$.

\begin{acknowledgments}
JMZP acknowledges financial support from the PCI program of the Brazilian agency ``Conselho Nacional de Desenvolvimento Científico e Tecnológico'' -- CNPq. SEJ thanks FAPERJ for the financial support. RRRR thanks CNPq for partial financial support (grant no. $309868/2021-1$). SBD thanks CNPq for partial financial support. This work has been done as a part of the Project INCT-Física Nuclear e Aplicações, Project number $464898/2014-5$. JDVA thanks Universidad Privada del Norte and Universidad Nacional Mayor de San Marcos for the financial support - RR Nº$\,005753$-$2021$-R$/$UNMSM under the project number B$21131781$.
\end{acknowledgments}

\end{document}